# Negative Longitudinal Magneto-Thermoelectric Power in a Semiconductor Parabolic Quantum Well

Firudin M. HASHIMZADE[*], Mirbaba M. BABAYEV, and Khanlar A. HASANOV

*Institute of Physics, National Academy of Sciences of Azerbaijan, Baku AZ 1143, Azerbaijan*

We present a theoretical study of the electronic thermoelectric power of a semiconductor parabolic quantum well in a magnetic field. The case of a longitudinal magnetic field, with respect to the temperature gradient, has been considered. The calculations were carried out taking into account spin-splitting of the dimensionally quantized electronic energy levels. It has been shown that in the region of strong confinement the thermoelectric power decreases with increasing magnetic field, which is related to the downward shift of the lower Zeeman-split spin subband.

---

[*] E-mail address: hashimzade@physics.ab.az

## 1. Introduction

The transport phenomena in low-dimensional systems presently attract a very high interest. The majority of such research is focused on the electrical properties. A significant number of papers are devoted to the thermoelectric phenomena in two-dimensional quantum well structures and wires.[1-17] In the paper by Bloch, the electronic thermoelectric power in a dimensionally quantized semi-metallic wire or in a degenerated semiconductor in a strong magnetic field has been studied theoretically.[1] The cases of longitudinal and transverse with respect to the wire magnetic fields have been considered. It has been shown that thermoelectric power oscillates with the width of the wire and the magnitude of the magnetic field. In the papers by Kubakaddi *et al.*[2] and Jali *et al.*[3] the expressions for the thermoelectric power in a two-dimensional quantum well structure and in a quantum wire have been derived for different mechanisms of electronic scattering. Beenakker and Staring developed a theory of thermoelectric power in a quantum dot.[4] Lyo calculated the thermoelectric power in a layered superlattice in a strong magnetic field.[5] Ghatak investigated the electronic thermoelectric power in a quantum wire of non-parabolic semiconductor in crossed electric and magnetic fields.[6,7] Ghatak and Mondal showed that thermoelectric power in a quantum film grows with increasing thickness of the film and with increasing magnetic field.[8]

After the prediction by Hick and Dresselhaus regarding a significant increase in the thermoelectric figure of merit for low-dimensional systems an interest to theoretical investigation of thermoelectric properties of these systems surged.[9,10] According to these authors, if a bulk material has good thermoelectric properties then the two-dimensional, and even more so, the one-dimensional material must have even better thermoelectric properties. Quantum confinement leads to an increase in the density of states in the vicinity of Fermi level, and, as a result, - to an increase in Seebeck coefficient. A theoretical study of this effect has been done for a bismuth nanowire.[11] Empirically the longitudinal thermoelectric effect as a function of the magnetic field was studied by Heremans and Thrush.[12] Later, the theory of thermoelectric power in superlattices of quantum wells and quantum wires was developed by Broido and Reinecke.[13] Fletcher *et. al.* showed that thermoelectric power carried information about the interaction between the two-dimensional electronic gas and the phonons, and at the low temperatures the relaxation time due to electron-phonon interaction can be determined with high precision from the measurements of the thermoelectric power.[14] The transverse thermoelectric power of a semiconductor film with parabolic confining potential was calculated by Gashimzade *et al.*[15] Mechanism of the growth of thermoelectric power in multiple quantum well has been studied theoretically for n-PbTe/PtEuTe by Koga *et al.*[16] A semi-classical theory of magneto-thermoelectric power of a quasi-two-dimensional electronic gas was developed by Zianni *et al.*[17]

None of the papers cited above took into account the effect of Zeeman splitting of dimensionally quantized levels upon thermoelectric power.

In the present paper we study the thermoelectric power in a quantum well with parabolic potential in a longitudinal magnetic field, taking into account Zeeman splitting. In general, the presence of the magnetic field leads to the increase in thermoelectric power. However, as it has been mentioned by Askerov et al.[18] in bulk materials for a certain region of quantizing magnetic fields a negative magneto-thermoelectric power can be observed.

## 2. Calculation of the Thermoelectric Power

Let us consider an electronic gas in a quantum well with the potential $U = \dfrac{m\omega_0^2 x^2}{2}$ in an external magnetic field ($B$) with the temperature gradient parallel to the magnetic field: $\nabla T \parallel \vec{B} \parallel Oz$. Here $m$ is the effective mass of electrons and $\omega_0$ is the parameter of the parabolic potential. The $x$ axis is perpendicular to the layer of electron gas. Choosing the Landau calibration for the vector potential as $A(0, x \cdot B, 0)$ one can present the law of dispersion of electrons as:

$$\varepsilon_{N,k_y,k_z,\sigma} = \left(N + \frac{1}{2}\right)\eta\omega + \frac{\eta^2 k_z^2}{2m} + \frac{\omega_0^2}{\omega^2}\frac{\eta^2 k_y^2}{2m} + \sigma g \mu_B B, \tag{1}$$

where $\omega = \sqrt{\omega_0^2 + \omega_c^2}$, $\omega_c = \dfrac{eB}{mc}$ is the cyclotron frequency of electrons, $(-e)$ is the charge of an electron, $c$ is the light speed, $g$ is the spectroscopic splitting factor, $\sigma = \pm\dfrac{1}{2}$ is the spin quantum number, $N$ is the oscillation quantum number, $\mu_B = \dfrac{e\eta}{2m_0 c}$ is the Bohr magneton, and $m_0$ is the mass of the free electron.[19,20]

Because the magnetic field does not affect the motion of the electrons along $z$-axis, for the distribution function of electrons $f(N,\sigma,k_y,k_z)$, we can use Boltzmann's equation.[1] Then for the current density we can write:

$$j_z = \sum_{N,k_y,k_z,\sigma}(-e) f(N,\sigma,k_y,k_z) v_z. \tag{2}$$

Here $v_z = \dfrac{\eta k_z}{m}$ is the velocity of an electron, $k_y, k_z$ is the components of the wave vector of the electrons. Using the relaxation time approximation we can write:

$$f(N,\sigma,k_y,k_z) = f_0(\varepsilon) + \upsilon_z \tau(\varepsilon) \frac{\partial f_0}{\partial \varepsilon}(eE_z + \frac{\varepsilon-\zeta}{T}\nabla_z T), \tag{3}$$

where $\tau(\varepsilon)$ is the electron relaxation time, and $\zeta$ is the chemical potential of electrons.[21]

For the case of electronic scattering by acoustic phonons the relaxation time $\tau(\varepsilon)$ can be written as $\tau(\varepsilon) = \tau_0 g^{-1}(\varepsilon)$ where $g(\varepsilon)$ is the density of states of the electrons, and $\tau_0$ is a constant (independent of the electronic energy). It is true that the inverse relaxation time is proportional to the modulated density of states with form-factor depending on the initial ($N,k_y,k_z$) and final ($N_1,k_{y1},k_{z1}$) electron state. However, further we consider the quantum limit, i.e. $\frac{\eta\omega}{k_0 T} \gg 1$, and $\omega_c \ll \omega_0$, in which case the above formula is applicable.

The expression for the density of states is given by Hashimzade et al.[22] We are considering case where the average energy of electrons is small compared to the depth of the parabolic well. Then the density of states of the electrons is described by a step function of energy of electrons:

$$g(\varepsilon) = \frac{L_y L_z m}{2\pi \eta^2} \cdot \frac{\omega}{\omega_0} \sum_{N\sigma} H(\varepsilon - \varepsilon_{N\sigma}). \tag{4}$$

Where

$$\varepsilon_{N\sigma} = (N+\frac{1}{2})\eta\omega + \sigma g \mu_B B, \tag{5}$$

$L_y$ and $L_z$ are lengths along the $y$ and $z$ directions. For each value of $\varepsilon$, $g(\varepsilon)$ is a sawtooth function of the magnetic field $B$. In the same paper,[22] the expression for the surface density of electrons is given:

$$n = \frac{n_0}{2}\frac{\omega}{\omega_0}\sum_N \left\{\ln\left[e^{\eta-\left(N+\frac{1}{2}\right)a+\frac{b}{2}}+1\right] + \ln\left[e^{\eta-\left(N+\frac{1}{2}\right)a-\frac{b}{2}}+1\right]\right\}, \tag{6}$$

where

$$n_0 = \frac{mk_0 T}{\pi\eta^2}, \quad a = \frac{\eta\omega}{k_0 T}, \quad b = \frac{|g|\mu_B B}{k_0 T}, \quad \eta = \frac{\xi}{k_0 T}. \tag{7}$$

Thus,

$$\tau^{-1}(\varepsilon) = \tau_0^{-1}\frac{L_y L_z m}{2\pi\eta^2}\frac{\omega}{\omega_0}\sum_{N_1\sigma_1} H\left[\varepsilon - \left(N_1+\frac{1}{2}\right)\eta\omega - \sigma_1 g\mu_B B\right]. \tag{8}$$

We must note that the relaxation time for the case of electronic scattering at point defects (short-range potential) and at rough surfaces has the same form.

Upon substitution of (3) and (8) in (2), defining $k'_y = k_y \frac{\omega_0}{\omega}$, and integrating in polar coordinates we obtain

$$j_z = \frac{e\tau_0}{m} \sum_{N,\sigma} \int_{\varepsilon_{N,\sigma}}^{\infty} \frac{\left(\varepsilon - \left(N+\frac{1}{2}\right)\eta\omega - \sigma g\mu_B B\right)\left(eE_z + \frac{\varepsilon-\zeta}{T}\nabla_z T\right)}{\sum_{N_1} H\left[\varepsilon - \left(N_1+\frac{1}{2}\right)\eta\omega - \sigma g\mu_B B\right]} \left(-\frac{\partial f_0}{\partial \varepsilon}\right) d\varepsilon. \qquad (9)$$

Dividing the area of integration in (9) as $\int_0^{\infty} \to \sum_{r=0}^{\infty} \int_{ra}^{(r+1)a}$ and integrating by parts, from setting the current density equal to zero we obtain the following expression for the thermoelectric power:

$$\alpha = -\frac{k_0}{e}\left(\frac{I_{20}+I_{11}}{I_{10}} - \frac{\zeta}{k_0 T}\right). \qquad (10)$$

Here, the following notations were used:

$$I_{10} = \sum_{\sigma=-1/2}^{1/2}\left[\sum_{N=1}^{\infty}\left(\frac{-a/2}{e^{\left(N+\frac{1}{2}\right)a+b\sigma-\eta}+1}\right) + Log\left(1+e^{\eta-\frac{a}{2}-b\sigma}\right)\right], \qquad (11)$$

$$I_{11} = \frac{1}{12}\sum_{\sigma=-1/2}^{1/2}\left\{\sum_{N=1}^{\infty}\left[\frac{(2N+1)a^2+6ab\sigma}{e^{\left(N+\frac{1}{2}\right)a+b\sigma-\eta}+1} - 6aLog\left(1+e^{\eta-\left(N+\frac{1}{2}\right)a-b\sigma}\right)\right] - \right.$$
$$\left. - 6(a+2b\sigma)Log\left(1+e^{\eta-\frac{a}{2}-b\sigma}\right)\right\}, \qquad (12)$$

$$I_{20} = \sum_{\sigma=-1/2}^{1/2}\left\{\sum_{N=1}^{\infty}\left[\frac{1}{6}\frac{(2N+1)a^2}{e^{\left(N+\frac{1}{2}\right)a+b\sigma-\eta}+1} + aLog\left(1+e^{\eta-\left(N+\frac{1}{2}\right)a-b\sigma}\right)\right] - \right.$$

$$-\left[Log\left(1+e^{\eta-\frac{a}{2}-b\sigma}\right)\right]^2 - 2PolyLog\left(2, \frac{1}{1+e^{-\eta+\frac{a}{2}+b\sigma}}\right)\right\}. \tag{13}$$

### 3. Non – Degenerate Electronic Gas

For a non-degenerate electronic gas the expression for the thermoelectric power can be considerably simplified:

$$\alpha = -\frac{k_0}{e}\left\{\frac{a(1+a)}{2+a-2e^a} + \frac{a}{e^a-1} + 2 + \frac{a}{2} - Log\left(\frac{Sinh\left(\frac{a}{2}\right)}{\frac{a}{2}}\right) - \right.$$

$$\left. -\frac{b}{2}Tanh\left(\frac{b}{2}\right) + Log\left(Cosh\left(\frac{b}{2}\right)\right) - Log\left(\frac{n}{n_1}\right)\right\}. \tag{14}$$

Here

$$n_1 = \frac{m(k_0 T)^2}{\pi \eta^3 \omega_0}. \tag{15}$$

One can see from the above that in the range of weak magnetic fields,

$$\frac{\Delta\alpha}{\alpha} = \frac{\frac{1}{2}\left(\frac{\omega_c}{\omega_0}\right)^2 - \frac{b^2}{8}}{2 - Log\left(\frac{n}{n_0}\right)}. \tag{16}$$

Thermoelectric power decreases with increasing magnetic field if $\frac{1}{2}\left(\frac{\omega_c}{\omega_0}\right)^2 < \frac{b^2}{8}$. This decrease is related to the downward shift of the bottom of the conductivity band due to the spin-splitting, contrary to the upward shift due to the dimensional-magnetic quantization. The latter prevails at sufficiently strong magnetic fields, in which case the thermoelectric power increases with increasing magnetic field.

### 4. Quantum Limit

This case is interesting because the negative magneto-thermoelectric power effect described above is the most pronounced in the quantum limit, at $a \gg 1$. In this case the expression for the thermoelectric power can be rewritten as

$$\alpha = -\frac{k_0}{e}\left\{\frac{\omega n_0}{n\omega_0}\left[PolyLog\left(2,\frac{\Phi(b)}{e^{\frac{b}{2}}+\Phi(b)}\right)+PolyLog\left(2,\frac{\Phi(b)}{e^{-\frac{b}{2}}+\Phi(b)}\right)+\right.\right.$$

$$\left.\left.+\frac{b}{4}Log\left(\frac{1+e^{-\frac{b}{2}}\Phi(b)}{1+e^{\frac{b}{2}}\Phi(b)}\right)\right]+\frac{2\omega_0 n}{\omega n_0}-Log(\Phi(b))\right\}, \quad (17)$$

where we introduced the notation

$$\Phi(b)=-Cosh\left(\frac{b}{2}\right)+\sqrt{-1+e^{\frac{2\omega_0 n}{\omega n_0}}+\left(Cosh\left(\frac{b}{2}\right)\right)^2}. \quad (18)$$

In Fig. 1 the dependence of the magneto-thermoelectric power with magnetic field is plotted. The numerical calculations were carried out for the following values of the parameters: $T=20K$, $g=50$, $m=0.016m_0$, $\eta\omega_0=20meV$, $n=n_0$. For these values the surface concentration is equal to $1.16\cdot 10^{10}cm^{-2}$. As one can see from Fig. 1, the relative decrease in thermoelectric power achieves 40% at the minimum, which is a significant change and can be easily detected in an experiment.

## 5. Classical Limit

This case corresponds to $a \ll 1$ and the replacement of summation over the quantum number $N$ by integration:

$$\alpha = -\frac{k_0}{e}\left(\frac{2(Log(1+e^\eta))^2+2PolyLog(2,-e^\eta)+4PolyLog\left(2,\frac{1}{1+e^{-\eta}}\right)}{Log(1+e^\eta)}-\eta\right). \quad (19)$$

Let us consider the particular cases of classical and degenerate statistics. Just as one should expect, we obtain well-known formulae for thermoelectric power in the absence of dimensional and magnetic quantization. At $\eta < 0$

$$Log(1+e^{\eta}) \approx e^{\eta}; PolyLog(2,-e^{\eta}) \approx -e^{\eta}; PolyLog\left(2,\frac{1}{1+e^{-\eta}}\right) \approx e^{\eta}. \qquad (20)$$

Therefore,

$$\alpha = -\frac{k_0}{e}(2-\eta). \qquad (21)$$

At $\eta \gg 1$

$$Log(1+e^{\eta}) \approx \eta; PolyLog(2,-e^{\eta}) \approx -\frac{\pi^2}{6} - \frac{1}{2}\eta^2; PolyLog\left(2,\frac{1}{1+e^{-\eta}}\right) \approx \frac{\pi^2}{6}, \qquad (22)$$

which implies

$$\alpha = -\frac{k_0}{e}\frac{\pi^2}{3\eta}. \qquad (23)$$

## 6. Conclusions

The electronic thermoelectric power have been calculated for a semiconductor quantum well with parabolic confining potential in an external magnetic field applied along the temperature gradient, parallel to the plane of free motion of electron. It has been shown that in a weak magnetic field thermoelectric power decreases with increasing magnetic field. This decrease is caused by the downward shift of the bottom of the conductivity band due to the spin-splitting, contrary to the upward shift due to the dimensional-magnetic quantization. The latter prevails at sufficiently strong magnetic fields, in which case the thermoelectric power increases with increasing magnetic field.

Fig.1. The dependence of the relative change of the magneto-thermoelectric power with magnetic field for a quantum well with $T = 20K$, $g = 50$, $m = 0.016m_0$, $\eta\omega_0 = 20meV$, $n = 1.16 \cdot 10^{10} cm^{-2}$.

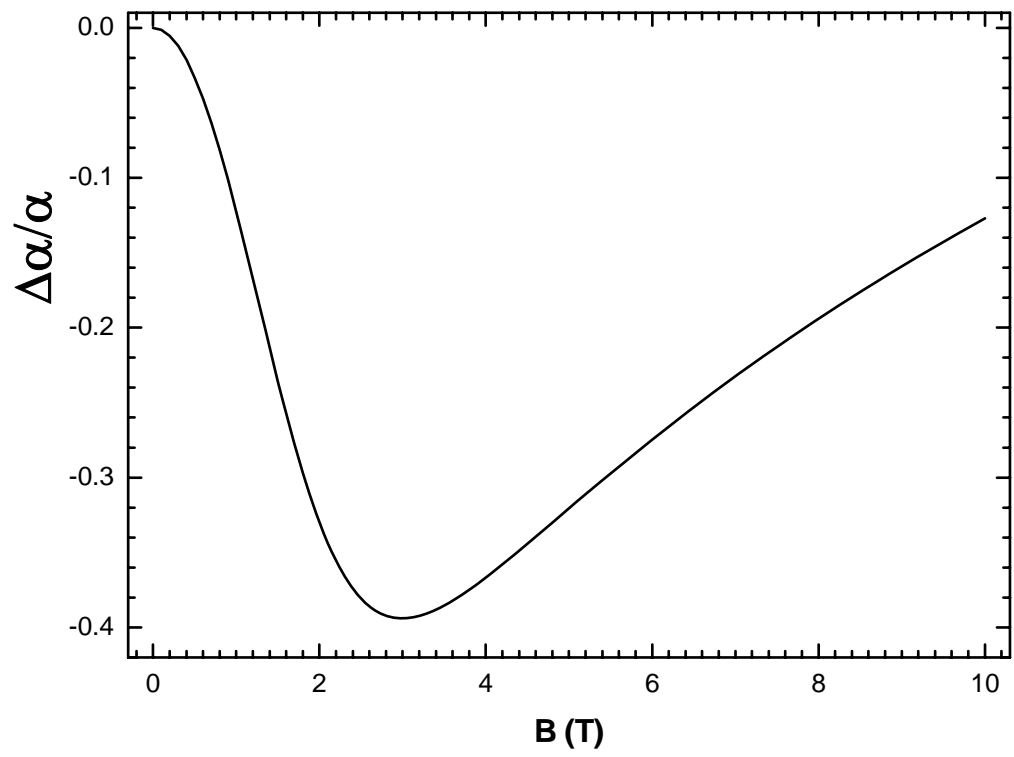